\documentclass[a4paper,12pt]{article}

\usepackage{amsfonts}
\usepackage{amsmath}
\usepackage{amsthm}
\usepackage{graphicx}
\usepackage{subfigure}

\usepackage{hyperref}

\begin{document}

\title{Statistical and dynamical decoupling of the IGM from Dark Matter}

\author{Li-Zhi Fang \\ Department of Physics, University of Arizona, \\
Tucson, AZ 85721, USA \\
fanglz@physics.arizona.edu\\
and\\
Weishan Zhu \\ Department of Physics, University of Arizona,\\
Tucson, AZ 85721, USA\\
Purple Mountain Observatory, Nanjing,210008, P.R. China \\
wszhu1985@gmail.com\\
}

\maketitle
\begin{abstract}

The mean mass densities of cosmic dark matter is larger than that of 
baryonic matter by a factor of about 5 in the $\Lambda$CDM universe.
Therefore, the gravity on large scales should be dominant by the distribution 
of dark matter in the universe. However, a series of observations incontrovertibly
show that the velocity and density fields of baryonic matter are decoupling from 
underlying dark matter field. This paper shows our attemps to unveil the physics 
behind this puzzle. In linear approximation, the dynamics of the baryon 
fluid is completely governed by the gravity of the dark matter. Consequently, the mass 
density field of baryon matter $\rho_b({\bf r},t)$ will be proportional to that of dark matter
$\rho_{\rm dm}({\bf r},t)$, even though they are different from each other initially.
In weak and moderate nonlinear regime, the dynamics of the baryon fluid can be
sketched by Burgers equation. A basic feature of the Burgers dynamics is to yield
shocks. When the Reynolds number is large, the Burgers fluid will be in the
state of Burgers turbulence, which consists of shocks and complex structures.
On the other hand, the collisionless dark matter may not show such shock, but
a multivalued velocity field. Therefore, the weak and moderate nonlinear evolution
leads to the IGM-dark matter deviation. Yet, the velocity field of Burgers fluid is 
still irrotational, as gravity is curl-free. In fully nonlinear regime, the vorticity 
of velocity field developed, and the cosmic baryonic fluid will no longer be potential, 
as the dynamics of vorticity is independent of gravity and can be self maintained by the 
nonlinearity of hydrodynamics. In this case, the cosmic baryon fluid is in the state of 
fully developed turbulence, which is statistically and dynamically decoupling from dark 
matter. This scenario provides a mechanism of cohenent explanation of observations.

\end{abstract}

\section{Introduction}

The mass field of the universe is dominated by dark matter, and only
a tiny fraction of cosmic matter is in the form of baryonic particles.
Gravitational clustering of dark matter is the major imputs of
cosmic structure formation. Collapsed massive halos of dark matter
are believed to be the hosts of various observed objects.
On the other hand, light-emitting and absorbing objects are made
up of baryonic matter.

Most of the baryonic matter in the universe actually is made of 
intergalactic gas, or intergalactic medium (IGM), mainly consisting
of neutral and ionized hydrogen, neutral and ionized helium \cite{bib1,bib2,bib3}. For a
long time, the mass field of the IGM, $\rho_{b}({\bf r})$, is
believed to be proportional to dark matter field $\rho_{\rm dm}({\bf
r})$. The similarity relation  $\rho_{b}({\bf r}) \propto \rho_{\rm
dm}({\bf r})$ should be hold, at least, on scales larger than the
thermal diffusion or Jeans length of the IGM. In other words, the
IGM mass field is considered as a point-by-point tracer of the dark
matter distribution. Obviously, in this case, the baryon fraction
$\rho_{b}({\bf r})/ [\rho_{\rm dm}({\bf r})+\rho_{b}({\bf r})]$ is
independent of ${\bf r}$, and equals to the cosmic mean
$\Omega_{b}/(\Omega_{b}+\Omega_{\rm dm})=0.17\pm0.01$ \cite{bib4,bib5}, 
where $\Omega_{b}$ and
$\Omega_{\rm dm}$ are, respectively, the cosmological parameters of
baryonic and dark matter

However, the similarity $\rho_{b}({\bf r}) \propto \rho_{\rm dm}({\bf
r})$, is found to be seriously inconsistent with observation in collapsed 
objects. The baryon fractions in gravitational bound objects generally are much lower
than the cosmic mean. It raises the so-called missing baryon problem.
The baryon fraction is claimed to decrease monotonically with the mass of
collapsed halos. For X-ray clusters and groups, the baryon fractions
are smaller than the cosmic mean by a factor of 2 - 4 \cite{bib6,bib7,bib8}.
The baryon fraction of dwarf galaxies can be as low as about $1\%$ of 
the cosmic mean \cite{bib9}.

Many works have been carried out  to explain the deficit
of baryon content in gravitational collapsed objects, including 
feedback due to supernovae \cite{bib10} or Active Galactic Nuclei(AGN) \cite{bib11,bib12}
, which blows out hot gas from halos, and pre-heating \cite{bib13,bib14}, which heats
baryon gas in the IGM and slows down the accrection of the IGM 
into halos. However, cosmologically hydrodynamic simulations
show that these mechanisms fail to produce the baryon
fraction in halos reveiled by observation \cite{bib15,bib16,bib17}

Actually, the baryon-dark matter decoupling probably is due mainly to 
the nonlinear nature of the hydrodynamics of the velocity and mass fields of 
IGM. That is, even without extra heating and cooling, the
hydrodynamical evolution can already provide mechanisms of separating 
the baryonic matter from dark matter. A possible
dynamical mechanism driving the baryonic matter deviate from dark matter
has been addressed in the pioneer work of Shandarin
\& Zeldovich \cite{bib18}. As the dark matter particles are assumed to be
collisionless, their velocities would be  multi-valued at the intersection
of trajectories. However, the baryon matter has a
single-valued velocity field as a compressible fluid. Discontinuities, such as shocks
or complex structures, will develop inevitably in the baryon fluid at these intersectiones.
Shocks would then cause the under-dense of baryonic matter in the post-shock area.
On the other hand, dark matter is not affected by the shocks. Thus, discontiunities
might prevent baryon matter to follow dark matter. Consequently, the baryon fraction in
the collapsed halos would be less than cosmic mean \cite{bib19}.

Therefore, the nonlinear evolution of IGM would be a major reason
of the statistical and dynamical decoupling between the cosmic baryon
fluid and dark matter. We report the results of this mechanism. This
paper will be arranged as follows. \S 2 is on the linear theory of
evolution of IGM perturbations. \S 3 presents the decoupling in
moderate nonlinear regime. \S 4 studies the fully nonlinear
development of the IGM. It shows the cosmic baryon fluid to be
turbulent on small scales. \S 5 gives observation evidences. The
conclusions will be in \S 6.

\section{Linear evolution of the perturbations of IGM mass field}

\subsection{Basic equations}

Let us consider a flat universe having cosmic factor $a(t)\propto
t^{2/3}$, and dominated by dark matter. We describe the IGM by a
mass density field $\rho({\bf x}, t)$ and a peculiar velocity field
${\bf v}({\bf x}, t)$, where ${\bf x}$ is the comoving coordinate.
In hydrodynamical description, the equations of the IGM consist of the
continuity and the momentum (or the Eular) equations as \cite{bib20}
\begin{equation}
\frac{\partial \delta}{\partial t} +
  \frac{1}{a}\nabla \cdot (1+\delta) {\bf v}=0
\end{equation}
\begin{equation}
\frac{\partial a{\bf v}}{\partial t}+
 ({\bf v}\cdot \nabla){\bf v}=
-\frac{1}{\rho}\nabla p - \nabla \phi
\end{equation}
where the density perturbation of the IGM $\delta({\bf x},t)=
[\rho{\bf x},t)-\bar{\rho}]/\bar{\rho}$, and $\bar{\rho}$ is the
mean density of the IGM. $p$ is the pressure of the IGM.

The IGM can be considered as a ''passive substance" with respect to
the underlying dark matter. Namely, the evolution of the IGM mass field
$\rho({\bf x},t)$ is governed by the gravity of total matter (baryonic
plus dark matter).  The peculiar gravitational potential
$\phi$ in eq.(2) satisfies
\begin{equation}
\nabla^2 \phi = 4\pi G a^2\bar{\rho}_t\delta_t.
\end{equation}
where the operator $\nabla$ is acting on the comoving coordinate
${\bf x}$, and $\delta_t({\bf x},t)=[\rho_t({\bf x}, t)
-\bar{\rho}_t]/\bar{\rho}_t$ is the perturbation of total density
field, with mean density $\bar{\rho}_t =1/6\pi Gt^2 \propto a^{-3}$.
The potential $\phi$ will be zero (or constant) when density
perturbation $\delta_t=0$.

To sketch the gravitational clustering of the IGM, we will not
consider the details of heating and cooling. Thermal processes
generally are local, and therefore, it is reasonable to describe the
thermal processes by a polytropic relation $p({\bf x},t) \propto
\rho^{\gamma}({\bf x},t)$. Thus eq.(2) can be rewritten as
\begin{equation}
\frac{\partial a{\bf v}}{\partial t}+
 ({\bf v}\cdot \nabla){\bf v}=
  -\frac{\gamma k_B T}{\mu m_p} \frac{\nabla \delta}{(1+\delta)}
  - \nabla \phi
\end{equation}
where the parameter $\mu$ is the mean molecular weight of
the IGM particles, and $m_p$ is the proton mass. In this case, we don't
need energy equation, and the IGM temperature evolves as $T \propto
\rho^{\gamma-1}$, or $T =T_0(1+\delta)^{\gamma-1}$. Exactly speaking,
there is also a noise in the hydrodynamical equations eq.(2) and
(4), which are related to the viscous of the IGM fluid due to the
fluctuation-dissipation theorem. Since the IGM is a ''passive
substance", this noise would be ignorable in searching for the
deviation of the IGM from the underlying dark matter.

\subsection{Linear solution of growth mode}

The unperturbed solutions of the density and velocity fields of
baryonic matter are $\bar{\rho}=[\Omega_b/(\Omega_b+\Omega_{\rm dm})]\bar{\rho}_t
\propto a^{-3}$, and ${\bf v}=0$. Therefore, the linearizion of
eqs. (1) and (4) yields
\begin{equation}
\frac{\partial \delta}{\partial t} +
  \frac{1}{a}\nabla \cdot {\bf v}=0
\end{equation}
\begin{equation}
\frac{\partial a{\bf v}}{\partial t} = -\frac{\gamma k_B
\bar{T}}{\mu m_p}\nabla \delta - \nabla \phi
\end{equation}
where the mean temperature $\bar{T}\propto \bar{\rho}^{\gamma-1}
\propto a^{-3(\gamma-1)}$. In the Fourier space, we have
\begin{equation}
\frac{\partial^2 \delta({\bf k},t)}{\partial t^2} + 2\frac
{\dot{a}}{a}\frac{\partial\delta({\bf k}, t)}{\partial t}+
\frac{1}{t^2}\frac{k^2}{k_J^2}\delta({\bf k},t) = 4\pi
G\bar{\rho}\delta_t({\bf k},t)
\end{equation}
\begin{equation}
\frac{\partial \upsilon({\bf k},t)}{\partial t}+
\frac{\dot{a}}{a}\upsilon({\bf k},t)= -
\frac{1}{t^2a}\frac{1}{k^2_J}\delta({\bf k},t)
  +\frac {4\pi G \bar{\rho}a}{k^2}\delta_t({\bf k},t),
\end{equation}
where ${\bf v}({\bf k},t)=i{\bf k}\upsilon({\bf k},t)$, and the
Jeans wavenumber $k_j$ is given by $k^2_J=(a^2/t^2)(\mu m_p/\gamma k_b \bar{T})
\propto a^{3\gamma -4}$. When $\gamma = 4/3$, $k_J$ is
time-independent. Here, we ignore the stochastic nature of
eqs.(7) and (8), and the randomness of all the variables are given
by the random initial conditions.

We consider only the growth mode of the perturbation of dark matter,
i.e. $\delta({\bf k},t) \propto a$. In the case of $\gamma=4/3$, the
solution of eqs.(7) and (8) is \cite{bib21}
\begin{equation}
\delta({\bf k},t) =\frac{\delta_t({\bf k},t)}{1+3k^2/2k_J^2} +
 c_1t^{-(1+\epsilon)/6}+c_2t^{-(1-\epsilon)/6}
\end{equation}
where $\epsilon=(1-4k^2/9k_J^2)^{1/2}$, and constants $c_1$ and
$c_2$ depend on the initial condition $\delta({\bf k},0)$ and
$\upsilon({\bf k},0)$. Therefore, regardless of the initial condition
of the IGM, after a long evolution we always have
\begin{equation}
\delta({\bf k},t) = \delta_t({\bf k},t), \hspace{5mm} {\rm if} \ k
\ll k_J.
\end{equation}
This means that the evolution of the IGM will completely follow dark
matter, and the initial difference between $\delta({\bf k},t)$ and
$\delta_t({\bf k},t)$ will be forgotten. Eq.(10) yields
\begin{equation}
P_b(k) =P_{\rm dm}(k), \hspace{1cm} {\rm if} \ k \ll k_J.
\end{equation}
The power spectrum of IGM field, $P_b(k)$ has to be the
same as that of dark matter $P_{\rm dm}(k)$.

The linear solutions of the IGM have also been studied by using
different assumptions of the IGM thermal processes other than that used in
eq.(4) \cite{bib22,bib23}. A common feature of these solutions is
\begin{equation}
\delta({\bf k},t) = (1+ {\rm decaying \ terms})\delta_t({\bf k},t)
  + {\rm decaying \ terms}
  \hspace{1cm} {\rm if} \ k \ll k_J.
\end{equation}
Only the decaying terms are affected by the initial conditions, and
therefore, the solutions  of eqs.(10) and (11) hold in general, 
regardless of the specific assumptions of the IGM thermal processes.

The physical explanation of the solutions (10) and (12) is
straightforward. The IGM is only a tiny component in the cosmic mass
field and its evolution is completely governed by the gravity of
dark matter in the linear regime. After the decaying of the initial 
conditions, the IGM
should follows the same trajectory as dark matter on scales larger
than the Jeans length.

In linear approximation, we also have the linear relation as
follows:
\begin{equation}
\delta({\bf x},t) =-f\nabla \cdot {\bf v}({\bf x}, t),
\end{equation}
where the parameter $f$ depends on dark matter models \cite{bib24}

\section{Moderate nonlinear regime of the IGM evolution}

\subsection{Burgers' equation for cosmic baryonic fluid}

In the moderate regime, we may use linear relation eq.(13) for
$\delta$ in eq.(4). We have then
\begin{equation}
\frac{\partial a{\bf v}}{\partial t}+
 ({\bf v}\cdot \nabla){\bf v}=
  - \frac{\nu}{a}\nabla^2 {\bf v}
  - \nabla \phi
\end{equation}
where the coefficient $\nu$ is given by
\begin{equation}
\nu=\frac{\gamma k_BT_0}{\mu m_p (d \ln D(t)/dt)},
\end{equation}
in which $D(t)$ describes the linear growth behavior. The term with
$\nu$ in eq.(14) describes a diffusion which is characterized by the 
Jeans length $k_J$.

To sketch the evolution of gravitational clustering, only the growth
modes are of interest. For the growth modes, the velocity field is
irrotational, and we can define a velocity potential by
\begin{equation}
{\bf v}=- \frac {1}{a}\nabla \varphi.
\end{equation}
Substituting eq.(16) into eq.(14), we have
\begin{equation}
\frac{\partial \varphi}{\partial t}- \frac{1}{2a^2}(\nabla
\varphi)^2 - \frac{\nu}{a^2}\nabla^2 \varphi =\phi.
\end{equation}
Equation (17) is the stochastic-force-driven Burgers' equations or
the KPZ equations \cite{bib25} . It is the simplest
nonlinear Langevin equation for growth modes \cite{bib26,bib27,bib28}.
This equation can
also be found with a coarse-graining approximation \cite{bib29}.

Thus, in moderate nonlinear regime, the cosmic baryonic fluid
is similar to Burgers' fluid. The second term on the l.h.s. results from
convection, and the third on the l.h.s. describes relaxation of the
clustering by diffusion. The term on the r.h.s represents the gravitational
potential of dark matter, which provides the initially random
perturbations. Clustered structures will develop from the initial
seeds via the competition of the convection and diffusion.

\subsection{IGM as a Burgers' fluid}

The Burgers' equation contains two scales: the dissipation
length or the Jeans length $1/k_J$ and the correlation length $r_c$
of the random force, $\phi$. The intensity of the random force
$\phi$ can be characterized by the density contrast of dark matter
$\delta_{\rm dm}$. Many theoretical works on Burgers fluids have been
done\cite{bib30,bib31,bib32,bib33,bib34}. They show that turbulence will develop
in Burger fluid when the following Reynolds number is large \cite{bib33,bib35} 
\begin{equation}
(k_Jr_c)^{2/3}\langle \delta_{\rm dm}^2\rangle^{1/3} > 1.
\end{equation}
This condition corresponds to the system in which the dark matter field  
has entered the non-linear regime.  In the linear regime, $\delta_{\rm dm}\ll
1$, and therefore, perturbations on scale $r_c > 1/k_J$ will not
cause Burgers' turbulence in the IGM. In the moderate nonlinear regime,
$\delta_{dm}\geq 1$, and Burgers' turbulence on scales larger the
Jeans length can develop. Since the dark matter field becomes
nonlinear first on small scales and then on large scales, one can
expect that Burgers' turbulence in the IGM develops from small to large
scales.

When Burgers turbulence develops in the IGM, its velocity and mass
density fields will be intermittent and contains
discontinuities, or shocks. The probability distribution function
(PDF) of velocity field is long tailed \cite{bib33}. Since dark
matter is not affected by Burgers turbulence, the IGM velocity field
will dynamically decouple from the dark matter field on scales larger
than the Jeans length once Burgers turbulence is developed. The intermittence
of the IGM mass field, as a result of the Burgers equation, has also been 
emphasized in \cite{bib23}. 

Jeans diffusion will also lead to the decoupling between the
velocity fields of the IGM ${\bf v}$ and dark matter ${\bf v}_{\rm dm}$.
However, the decoupling given by the Jeans diffusion is very
different from that results from the Burgers turbulence shocks. For the
former, the distribution ${\bf v}$ and $\Delta {\bf v}\equiv {\bf
v}({\bf x +r})- {\bf v}({\bf x-r})$ is symmetric with respect to the
transformation ${\bf v}_{\rm dm} \rightarrow -{\bf v}_{\rm dm}$, i.e., the
velocity PDF with $\Delta {\bf v}\cdot{\bf v}_{\rm dm}/|{\bf v}_{\rm dm}|>0$
(acceleration in the direction of ${\bf v}_{\rm dm}/|{\bf v}_{\rm dm}|$) is
the same as the PDF with $\Delta {\bf v}\cdot{\bf v}_{\rm dm}/|{\bf
v}_{\rm dm}|<0$ (deceleration in the direction of ${\bf v}_{\rm dm}/|{\bf
v}_{\rm dm}|$). For Burgers turbulence, the shocks consist of an
acceleration ramp followed by a rapid deceleration. Therefore, the
IGM is not symmetric between the sections accelerating and
decelerating. Since the acceleration is due to the gravity of dark
matter, the acceleration is generally in the direction of ${\bf
v}_{\rm dm}$. Thus, the IGM velocity field ${\bf v}$ is asymmetric with
respect to the transformation ${\bf v}_{\rm dm} \rightarrow -{\bf
v}_{\rm dm}$. The PDF of $\Delta {\bf v}$ with $\Delta {\bf v}\cdot{\bf
v}_{\rm dm}/|{\bf v}_{\rm dm}|>0$ will not be the same as the PDF of $\Delta
{\bf v}$ with $\Delta{\bf v}\cdot{\bf v}_{\rm dm}/|{\bf v}_{\rm dm}|<0$.

If $\nu =0$, Burgers equation (17) is scale free when the random
force $\phi$ is scale-free. In this case, the velocity field of the
Burgers fluid is self-similar and the PDF of the velocity difference
$\Delta {\bf v}$ is scale-invariant.

In summary, if the IGM can be described as a Burgers' fluid, its
velocity field should show the following properties: 1.) the
peculiar velocity of the IGM at a given point will generally be
lower than that of dark matter at the same point, 2.) the PDF of
$\Delta {\bf v}_r\cdot{\bf v}_{\rm dm}/|{\bf v}_{\rm dm}|>0$ will not be the
same as the PDF of $\Delta {\bf v}_r\cdot{\bf v}_{\rm dm}/|{\bf
v}_{\rm dm}|<0$, and 3.) The PDF of $\Delta {\bf v}_r$ is scaling. These
predictions have been proven by cosmologically hydrodynamic
simulation samples \cite{bib36}. 

\section{Fully developed nonlinear regime}

\subsection{Vorticity}

Burgers turbulence is qualitatively different from turbulence
described by the Navier-Stokes equations. The latter generally
consists of vortices on various scales, while the former is a
collection of shocks. These features arise because for growth modes
of the perturbations of velocity fields is potential, and the velocity 
field is irrotational. Moreover, from the
condition eq.(18), we see that the Burgers turbulence can only
develop in dense areas of dark matter $\delta_{\rm dm}>1$.

However, in fully developed nonlinear regime, the cosmic baryon
fluid will on longer be potential, and the turbulence of
Navier-Stokes fluid can also be developed in high as well as low
density regions. This point can be seen with the vorticity, which is
defined by $\omega_i=(1/2)(\partial_iv_j-\partial_j v_i)$ or
$\vec{\omega}=\nabla \times {\bf v}$, where $i=1,2,3$. The dynamical
equation of the vorticity $\vec{\omega}$ of cosmic baryon fluid can
be derived from Navier-Stokes equations as follows \cite{bib37}
\begin{equation}
\frac{D \vec {\omega}}{Dt}\equiv \partial_t {\vec \omega}
+\frac{1}{a}{\bf v}\cdot{\nabla}\vec{\omega}=\frac{1}{a}({\bf
S}\cdot {\vec\omega}-d {\vec\omega} +\frac{1}{\rho^2}\nabla \rho
\times\nabla p-\dot{a}\vec{\omega}),
\end{equation}
where $d=\partial_iv_i$ is the divergence of the velocity field.
Tensor ${\bf S}$ is called strain rate defined as
$S_{ij}=(1/2)(\partial_iv_j+\partial_jv_i)$, and the vector $[{\bf
S}\cdot {\vec\omega}]_i=S_{ij}\omega_j$.

A remarkable property of eq.(19) is the absence of the term of
gravity of mass fields, indicating that the gravitational field of dark
matter is not the sources of the vorticity. This point is obvious,
as gravity is curl-free in nature. In this sense, the vorticity is
''non-gravitational''.

In the linear regime of perturbations, both the terms ${\bf S}\cdot
{\vec\omega}-d {\vec\omega}$ and $\nabla \rho \times\nabla p$ would
be zero. Vorticity $\vec {\omega}$ will decay as $a^{-1}$, given
by the term $\dot{a}\vec{\omega}$. Therefore, as expected, in
the linear regime the velocity field of baryon matter is irrotational.
In other words, the linear evolution of the velocity field of baryon
fluid is fully governed by the gravity.

However, in the fully developed nonlinear regime, the vorticity will no longer be zero.
First, it has been shown that at the moderate nonlinear regime,
Burgers' fluid will inevitably lead to the development of shocks and
complex structures, which can yield non-zero term $\nabla \rho
\times\nabla p$ of eq.(19), being called baroclinity. Thus, the vorticity is
excited, regardless of the gravity of dark matter. Physically, once
multi-streaming has developed, complex structures, like curved
shocks, will lead to deviation of the direction of $\nabla p$ from
that of $\nabla \rho$. Consequently, the density pressure relation
cannot be simply given by an single-variable function equation as
$p=p(\rho)$ \cite{bib38} and the baroclinity term will no longer
be zero.

Once vorticity is initiated, eq.(19) shows that the vorticity of
baryon velocity field can be maintained by itself, i.e. the system
actually is nonlinearly self-excited and self-maintained. The divergence
$d=\partial_jv_j$ is generally negative in regions of clustering.
Therefore, the term $-d\omega$ of eq.(19) will lead to an
amplification of vorticity in overdense regions. On the other hand,
the vorticity will be stretched by the nonlinear term ${\bf S}\cdot
{\vec\omega}$. Therefore, vorticity can also develop in low mass
density region. This picture is very different from the Burgers' 
turbulence.

Vortices are considered as a fundamental ingredient of turbulence.
The fluctuations of the vorticity field is an important measurement
of the turbulence of fluid \cite{bib39,bib40}.
Therefore, the dynamics of vorticity provides a non-gravitational
mechanism of the decoupling between baryon matter from
dark matter.

The turbulence of cosmic baryon fluid can also be seen with the
spectrum tensors $\Phi_{ij}({\bf k})$ and $\Omega_{ij}({\bf k})$,
which are, respectively, the Fourier counterparts of the two-point
correlation tensors of velocity $\langle v_i({\bf x+r})v_j({\bf
x})\rangle$ and vorticity $\langle \omega_i({\bf x+r})\omega_j({\bf
x})\rangle$. That is
\begin{equation}
\Phi_{ij}({\bf k})=\frac {1}{(2\pi)^3}\int \langle v_i({\bf
x+r})v_j({\bf x})\rangle e^{-i{\bf k}\cdot {\bf r}}d{\bf r}
\end{equation}
\begin{equation}
\Omega_{ij}({\bf k})=\frac {1}{(2\pi)^3}\int \langle \omega_i({\bf
x+r})\omega_j({\bf x})\rangle e^{-i{\bf k}\cdot {\bf r}}d{\bf r}.
\end{equation}
respectively, where $\langle ...\rangle$ denotes  average over
spatial coordinates ${\bf x}$. For a homogeneous turbulence, we have \cite{bib39}

\begin{equation}
\Omega_{ij}({\bf k})=[\delta_{ij}k^2-k_ik_j]\Phi_{ll} ({\bf
k})-k^2\Phi_{ij}({\bf k}),
\end{equation}
and hence,
\begin{equation}
\Omega_{ii}({\bf k})=k^2\Phi_{ii}({\bf k}).
\end{equation}
The power spectra of velocity and vorticity fields are defined
respectively as
\begin{equation}
P_v(k)=\int \frac{1}{2}\Phi_{ii}({\bf k})\delta(|{\bf k}|-k)d{\bf
k}; \hspace{3mm} P_{\omega}(k)=\int \frac{1}{2}\Omega_{ii}({\bf
k})\delta(|{\bf k}|-k)d{\bf k}.
\end{equation}
Combining eqs. (23) and (24) yields
\begin{equation}
P_{\omega}(k)=k^2P_v(k).
\end{equation}
This relation can be used to measure the developed level of
turbulence. If the velocity and vorticity fields of a fluid satisfy
the relation given by eq.(25), it should be in the state of fully
developed homogeneous turbulence. Otherwise, it would be less
developed. The samples of cosmological hydrodynamic simulation have
showed that the relation eq.(25) is perfectly hold on scales less
than $3 h^{-1}$ Mpc at $z\sim 0$ \cite{bib37}. Actually, the
dynamics of eq.(19) reveals that the cosmic baryonic fluid underwent
a scale-free evolution, which leads to a state of fully developed
turbulence.

\subsection{She-Leveque intermittence}

The velocity field of a fully developed turbulence is highly
intermittent, which is measured by the structure functions $S_p(r)$
defined as
\begin{equation}
S_p(r)\equiv \langle \delta v_r^p\rangle \sim r^{\zeta_p}.
\end{equation}
where $\delta v_{r}\equiv [({\bf v(x+r)- v(x)})\cdot {\bf r}/r]$ and
$\zeta_p$ is called intermittent exponent. Based on dimensional
argument of hierarchical evolution, Kolmogorov in 1941 \cite{bib41} predicted
that for fully developed turbulence, the intermittent exponent is $\zeta_p=p/3$ 
on scales of inertial range. Experimental and numerical
results do not, however, support the $p/3$ law. It should be
attributed to intermittency, i.e. turbulence field in inertial range
is characterized by stronger non-Gaussianity on smaller scales. A
remarkable development was made by She and Leveque (SL hereafter) \cite{bib42}
. They proposed that the non-Gaussian behavior of fully
developed turbulence is determined by the hierarchical structure
originated from the Navier-Stokes equation, and the $p/3$ law should
be replaced by
\begin{equation}
\zeta_p/\zeta_3=[1-C(1-\beta^3)]p/3+ C(1-\beta^{p}),
\end{equation}
which contains only two dimensionless parameter $C$ and $\beta$. $C$
is the Hausdorff dimension of the most dissipative structures, and
parameter $\beta$ is given by the hierarchical evolution. SL formula
eq.(27) is in excellent agreement with various experiments of
turbulence, including turbulence in compressible fluid. The SL
scaling law of structure function is considered to be universal for
characterizing the fully developed turbulence.

With the hydrodynamic simulation sample of the concordance
$\Lambda$CDM universe, the intermittent exponent of the velocity
field of cosmic baryon fluid at redshift $z=0$ in the scale range
from the Jeans length to about 16 h$^{-1}$ Mpc is found to be
extremely well described by the She-Leveque's universal scaling
formula eq.(27) with $C=1$ and $\beta^3=1/3$ \cite{bib43}. It
indicates that the dissipative structures are dominated by sheets.
These results imply once again that the evolution of highly evolved
cosmic baryon fluid is similar to a fully developed turbulence.

When a fluid is turbulent, the kinetic energy passes from large to the
smallest eddies, and finally dissipates into thermal motion. For
cosmic baryonic gas, the evolution of vorticity is also
hierarchical. The vorticity evolves from large scales to small
scales, and finally falls into massive halos to form structures,
including light-emitting objects. This result strongly indicates
that the highly nonlinear evolution would lead to the cosmic baryon
fluid reaching a statistically quasi-equilibrium state satisfying
the universal scaling as that of fully developed turbulence.

In view of this picture, we can say that in the highly nonlinear
regime, the statistical properties, especially the intermittent
behavior, of the velocity fluctuations are actually independent of
the details of the dissipative processes. The state depends only on
the dimension of dissipative structures and the hierarchical
relation index.

\section{Observable effects}

The nonlinearly evoloved IGM provides a coherent and uniform expanation of various observations, 
including the intermittence and non-Gaussianity of Ly$\alpha$ transmitted fllux of 
quasar absorption espectrum \cite{bib43a,bib43b}; Ly$\alpha$ leaks of absorption espectrum 
of high redshift objects\cite{bib43c}; the turbulence broadening\cite{bib43d};  Scaling 
relations between SZ effect and X-ray luminosity of X-ray clusters \cite{bib38,bib43e,bib43f}, 
baryon missing in collapsed halos \cite{bib19}; etc. Here we address only two of them.

\subsection{Non-Gaussianity of Log-Poisson hierarchy}

It has been shown that the SL scaling is yielded from the
Log-Poisson hierarchy process, which is related to the so-called
generalized scale covariance of the Navier-Stokes equations \cite{bib44}
. Therefore, the statistical behavior of the mass
field of cosmic baryon matter should be given by the log-Poisson
random multiplicative processes (RMP).

The log-Poisson RMP assumes that, in the scale-free range, the
variables $\delta\rho_{r}\equiv \rho({\bf x+r})-\rho({\bf x})$,
$r=|\bf r|$, on different scales $r$ are related from each other by
a statistically hierarchy relation as
\begin{equation}
|\delta\rho_{r_2}| = W_{r_1r_2}|\delta\rho_{r_1}|,
\end{equation}
where
\begin{equation}
W_{r_1r_2}=\beta^m (r_1/r_2)^{\gamma},
\end{equation}
which describes how the fluctuation $|\delta\rho_{r_1}|$ on the
larger scale $r_1$ related to fluctuations $|\delta\rho_{r_2}|$ on
the smaller scale $r_2$. In eq.(29), $m$ is a Poisson random
variable with the PDF
\begin{equation}
P(m)=\exp(-\lambda_{r_1r_2})\lambda_{r_1r_2}^m/m!.
\end{equation}
To insure the normalization $\langle W_{r_1r_2} \rangle=1$, where
$\langle...\rangle$ is over $m$, the mean $\lambda_{r_1r_2}$ of the
Poisson distribution should be
\begin{equation}
\lambda_{r_1r_2}= \gamma[\ln(r_1/r_2)]/(1-\beta).
\end{equation}
It is enough to consider only $|\delta\rho_{r}|$, as the
distribution of positive and negative $\delta\rho_{r}$ is symmetric.

The log-Poisson hierarchy given by eq.(29) depends only on the ratio
$r_1/r_2$, which is obviously scale invariant. The hierarchy is
determined by two dimensionless positive parameters: $\beta$ and
$\gamma$, describing, respectively, the intermittence and
singularity of the random fields. Equation (29) relates
$\delta\rho_{r}$ on different scales by multiplying a random factor
$W$, which generally yields a non-Gaussian field even if the field
originally is Gaussian \cite{bib45}.

The cascade from scale $r_1$ to $r_2$, and then to $r_3$ is
identical to the cascade from $r_1$ to $r_3$. It is because
$W_{r_1r_3}=W_{r_1r_2}W_{r_2r_3}=\beta^N (r_1/r_3)^{\gamma}$, where
$N$ is again a Poisson random variable with
$\lambda_{r_1r_3}=\lambda_{r_1r_2}+\lambda_{r_2r_3}=\gamma[\ln(r_1/r_3)]/(1-\beta)$.
Therefore, the log-Poisson hierarchy removes an arbitrariness in
defining the steps of cascade from $r_1$ to $r_2$ or $r_2$ to $r_3$.
The log-Poisson hierarchy is discrete in terms of the discrete
random number $m$. However, the scale $r$ is infinitely divisible.
Namely, there is no lower limit on the difference $r_1-r_2$. It can
be infinitesimal, and the hierarchical process is of infinite
divisibility.

With the log-Poisson hierarchy eqs.(28)-(31), the density field
should have the non-Gaussianities, which can be described by the
structure function of density field  $S_p(r)\equiv \langle
|\delta\rho_r|^p\rangle$ as follows \cite{bib46}:

1. Intermittent exponent. The structure function should be a
function of power law of $r$ as
\begin{equation}
S_p(r)\propto r^{\xi(p)},
\end{equation}
and the intermittent exponent $\xi(p)$ is given by
\begin{equation}
\xi(p)=-\gamma[p-(1-\beta^{p})/(1-\beta)].
\end{equation}
This is actually the SL scaling formula for density field.

2. Hierarchical relation. Defining
\begin{equation}
F_p(r) \equiv S_{p+1}(r)/S_p(r),
\end{equation}
the log-Poisson RMP gives hierarchical relation as 
\begin{equation}
\frac{F_{p}(r)}{F_{\infty}(r)}= \left
[\frac{F_{p+1}(r)}{F_{\infty}(r)}\right]^{1/\beta}.
\end{equation}
Relation (35) is invariant with respect to a translation in $p$.

3. Moments. The ratio between high order and 2nd order moments,
$\langle\delta\rho_r^{2p}\rangle/\langle\delta \rho_r^2\rangle^p$
should satisfy
\begin{equation}
\ln \frac{\langle \delta\rho_r^{2p}\rangle}{\langle
\delta\rho_r^{2}\rangle^p}= K_p\ln r + {\rm const}
\end{equation}
with
\begin{equation}
K_p=-\gamma\frac{p(1-\beta^2)-(1-\beta^{2p})}{1-\beta}.
\end{equation}

4. Scale-scale correlation. The so-called scale-scale correlation is
defined as
\begin{equation}
C^{p,p}_{r_1,r_2}\equiv \frac{\langle \delta \rho_{r_1}^{p}\delta
\rho_{r_2}^{p}\rangle} {\langle \delta \rho_{r_1}^{p}\rangle
\langle\delta \rho_{r_2}^{p}\rangle}.
\end{equation}
Obviously, for a Gaussian field, $C^{p,p}_{r_1,r_2}=1$. If the ratio
$r_2/r_1$ is fixed, the log-Poisson model predicts the scale-scale
correlation to be
\begin{equation}
C^{p,p}_{r_1,r_2}=B(r_2/r_1)r_1^{\xi(2p)-2\xi(p)},
\end{equation}
where the coefficient $B(r_2/r_1)$ depends only on the ratio
$r_2/r_1$, as the log-Poisson model is invariant of the dilation.

It should be emphasized again that all above-mentioned
non-Gaussianities depend only on two dimensionless parameters
$\beta$ and $\gamma$. It has been revealed that the fields of Ly$\alpha$
transmitted flux fluctuations of observed high resolution and high S/N Ly$\alpha$
absorption spectra of quasars at moderate redshift $z\sim$ 2 - 3, and 
high redshift $z\sim$ 5 - 6 do show all the log-Poisson non-Gaussianities 
\cite{bib47,bib47b}.

\subsection{Turbulence pressure and baryon missing}

As shown in \S 2.2, the Jeans length $\lambda_J=2\pi/k_J$ is
$\lambda_J=c_s\sqrt{\pi/G\rho}$, $c_s$ being the sound speed. The
effect of turbulent motions on the Jeans length can be estimated by replacing $c_s$
by an effective sound speed
\begin{equation}
c^2_{\rm s,eff}=c_s^2+\frac{1}{3} \langle v^2\rangle
\end{equation}
where $\langle v^2\rangle$ is the rms velocity dispersion due to
turbulent motion \cite{bib48,bib49}. Namely, the random velocity
field of turbulence contributes to an extra pressure, the turbulence
pressure, $p_{\rm tub}=\rho\langle v^2\rangle$ to resist the
gravitational collapsing.

If the collapsed halo is on scale $R$, the gravitational collapsing
will not be affected by the rms velocity dispersion on scales larger
than $R$, the fluctuation of velocity on the scales $k< 2\pi /R$ do
not contribute to the turbulent pressure in terms of resist
gravitational collapsing on scales that larger than $R$. Thus, the
effect of turbulent pressure on gravitational clustering should be
given by \cite{bib50}
\begin{equation}
p_{\rm tur}(k_{R})=\int_{k_{R}}^{k_{\rm max}} E(k) dk,
\end{equation}
where $E(k)$ is the power spectrum of kinetic energy of the
turbulence, and $k_{R}=2\pi/R$. Wavenumber ${k_{\rm max}=2\pi/l_{\rm
diss}}$ corresponds to the minimal scale $l_{\rm diss}$ below which
the turbulence decays due to energy dissipation or virialization.
The turbulent pressure is dynamical, not thermal. It can
be as strong as a thermal pressure of temperature $10^{5-6}$ K,
while the temperature of the IGM is still remained in the range
$10^{4-5}$ K.

Obviously, the bulk velocity will not contribute to the $\langle v^2\rangle$, as
it can be given by the motion of falling into a gravitational well.
Therefore, we should pick up the component of velocity field that
is regardless of gravity. It is mainly given by the vorticity of
velocity field. In other words, the turbulent vorticity field can prevent
the collapsing of baryon matter into the halos of dark matter. It
leads to the deviation of the mass density fields of baryon matter
from that of dark matter. 

With the samples produced by cosmological hydrodynamical simulation,
we find that the turbulence pressure will play an
important role of preventing the IGM clustering. In the IGM regions
with 10 times mean overdensity, the turbulent pressure is equivalent
to the thermal pressure of the baryon gas with temperature of about
$1\times 10^5$ K. The turbulent pressure is dynamical and
non-thermal. When the turbulence dissipated, the kinetic energy of
turbulence becomes the thermal energy. It yields the entropy in
halos \cite{bib38}. 

The baryon fraction of dark matter halos is found to be a strong function 
of the halo mass $M_{200}$, the mass within the range of total mass density 
larger than $200\bar{\rho}_t$. The
baryon fraction is significantly decreasing with the mass of dark
halos. For halos with mass $M_{200}= 3 \times 10^{10}$ M$_{\odot}$,
the baryon fraction is only about 0.1 of the cosmic mean of the
concordance $\Lambda$CDM universe, and it further decreases to 0.03
for $M_{200} = 10^{10}$ M$_{\odot}$ halos. This result is well
consistent with observed baryon content of galaxies and galaxy
groups. Therefore, the turbulent velocity field of baryon fluid
would be the major factor leading to the absence of baryon matter in
halos with mass less than $10^{11.5}$ M$_{\odot}$. With
this evolution picture, no pre-heating or reheating mechanisms are
needed for explaining baryon missing. 

\section{Conclusion}

The nonlinear evolution of cosmic baryon fluid is a central problem
of cosmology. The concordance $\Lambda$CDM model
provides an effective paradigm of formation and evolution of cosmic
structures. The dynamical behavior of the CDM component has been
successfully revealed with analytical study and N-body numerical
simulation. On the other hand, the dynamics of baryon fluid is less
understood, especially in the nonlinear regime, because the
involvement of complicated physical processes, such as cooling,
heating, ionization, shock, turbulence etc. On the other hand, many
observations are directly related to the nonlinear evolution of
baryon fluid, such as the absorption spectra of quasars, galaxies
and GRB at high redshift, Sunyaev-Zeldovich effect, the polarization
of cosmic microwave background (CMB), X-ray background etc. One of
the biggest challenges in galaxy formation is also on the physical
origin of the non-linear relation between baryon gas and low-mass
dark halos.

Thanks for the development of reliable cosmological hydrodynamical
simulation in the last decade, significant progress has been made
in understanding the non-linear dynamics of cosmic baryon fluid. The
turbulence behavior of cosmic baryon fluid in scale free regime is
the key of understanding nonlinear evolution of the mass density and
velocity fields. An important consequence of the turbulence is to
yield statistical decoupling of the mass and velocity fields of
baryon fluid from the underlying dark matter field. This result is
extremely valuable to understand the low baryon fraction in galaxy
clusters, the non-Gaussian features of quasar Ly$\alpha$ transmitted
flux, the Ly$\alpha$ leaking of absorption spectra of high redshift
objects, the shock heating of pre-virialized structures, and the
strongly nonlinear dependence of the star formation efficiency on
halo mass etc. Considering these complexities, the nonlinear
dynamics of the IGM also is essential for understanding the epoch of
reionization and related observations.



\end{document}